\documentclass[10pt,a4paper]{article}
\usepackage[dvips]{color}
\usepackage{epsfig}
\usepackage{epstopdf}
\usepackage{amsmath}
\usepackage{graphicx}
\usepackage{authblk}
\usepackage{amssymb,amsmath}
\usepackage{amsmath}
\usepackage{tabularx}

\begin{document}
\title{\bf Phase space analysis of some interacting Chaplygin gas models}
\author{{M. Khurshudyan$^{a}$\thanks{Email:khurshudyan@yandex.ru}~~and R. Myrzakulov$^{b}$\thanks{Email:rmyrzakulov@gmail.com}}\\
$^{a}${\small {\em Armenian State Pedagogical University, 375010 Yerevan, Republic of Armenia}}\\
$^{b}${\small {\em Eurasian International Center for Theoretical Physics, Eurasian National University, Astana 010008, Kazakhstan}}\\ }\maketitle

\begin{abstract}
The goal of this paper is to discuss phase space analysis of some interacting Chaplygin gas models in General Relativity. Chaplygin gas is one of the fluids actively considered in modern cosmology due to the fact that it is a joint model of dark energy and dark matter. In this paper we have considered various forms of interaction term $Q$ including linear and non linear sign changeable interactions. For each case late time attractors for the field equations are found.
\end{abstract}

\section{Introduction}\label{sec:INT}
One of the active topics of modern cosmology it is an accelerated expansion of the large scale universe~\cite{Riess}~-~\cite{Verde}. There are various approaches to solve the problem including dark energy~\cite{Jaewon}, different modifications of General Relativity~\cite{Timothy} and other interesting ideas~\cite{Benoit}~-~\cite{Buchert}~(to mention a few). There is an active research towards to all directions. If we keep General Relativity as a main theory of Gravity, then to describe dark energy we should assume either an explicit form of the EoS or the form of the energy density. It is not excluded, that the EoS of the dark energy can be a solution of an algebraic or a differential equation~\cite{Bamba},~\cite{Nojiri}, which could open a window towards fundamental description of darkness of the universe.  One of the models actively studied in modern cosmology it is Chaplygin gas~\cite{Kamenshchik},~\cite{Bento}, which has EoS of the following form
\begin{equation}\label{eq:CH}
P_{c} = A\rho_{c} - \frac{B}{\rho_{c}^{\alpha}},
\end{equation}
where $A$, $B$ and $\alpha$ are positive constants and $\rho_{c}$ it is the energy density of the gas. This is an example of the fluid which has an explicit EoS. It is well know that this fluid is a joint model of the dark energy and dark matter. Therefore, this is one of the main reasons to continue research on this model, despite to some critics~\cite{Havard}. From Eq.~(\ref{eq:CH}) the case of $A = 0$ recovers generalized Chaplygin gas EoS, and $A = 0$ together $\alpha = 1$ recovers the original Chaplygin gas EoS. The best fitted parameters are found to be $A = 0.085$ and $\alpha = 1.724$, while Constitution + CMB + BAO and Union + CMB + BAO results are $A = 0.061 \pm 0.079$, $\alpha = 0.053\pm0.089$, and $A = 0.110\pm0.097$, $\alpha = 0.089 \pm 0.099$ respectively~\cite{Xu},~\cite{Toribio}. Other observational constraints on modified Chaplygin gas model using Markov Chain Monte Carlo approach found that $A = 0.00189^{+0.00583}_{-0.00756}$, $\alpha = 0.1079^{+0.3397}_{-0.2539}$ at $1\sigma$ level and $A = 0.00189^{+0.00660}_{-0.00915}$ with $\alpha = 0.1079^{+0.4678}_{-0.2911}$ at $2\sigma$ level~\cite{Lu_Xu}. In recent Physical Literature different modifications of EoS of Chaplygin gas could be found~(see for instance~\cite{Sadeghi}~-~\cite{Yang} and references therein). On the other hand, dark energy does give to an origin of a problem known as cosmological coincidence problem~\cite{Velten}. In this problem we should explain why 
\begin{equation}
r=\frac{\Omega_{DM}}{\Omega_{DE}} = r_{0},
\end{equation}
where $r_{0}$ is a constant. To solve this problem an interaction between the components of the universe has been studied. An active research on interacting cosmological models reveals that interaction term $Q$ can change the sing during the evolution of the universe~\cite{Su},~\cite{Wei0}. While before it was thought that it is not possible. In this paper we will consider cosmological models of interacting Chaplygin gas, where dark matter is a pressureless fluid. Our goal is to consider various forms of interaction term $Q$ and to obtain late time attractors of the field equations describing the dynamics of such universe. Considered interactions are examples obtained from the forms presented below
\begin{equation}\label{eq:Q1}
Q = q^{n}(3Hb \rho + \gamma \dot{\rho} ),
\end{equation}
and
\begin{equation}\label{eq:Q2}
Q = 3 H b q^{n} \left( \rho + \frac{\rho_{i}\rho_{j}}{\rho } \right ),
\end{equation}  
where $b$, $\gamma$, and $n$ are positive constants, $H$ it is the Hubble parameter, $q$ it is the deceleration parameter, while $\rho$ can be either the energy density of the effective fluid or the energy density of one of the components. $\rho_{i}$ is the energy density of one of the components. We restrict our attention to the models where $n=0$ and $n=1$. Interactions described via Eq.~(\ref{eq:Q1}) and Eq.~(\ref{eq:Q2}) for $n=0$ have fixed sign during the whole evolution of the universe. For $n=1$ interactions Eq.~(\ref{eq:Q1}) and Eq.~(\ref{eq:Q2}) represent sign changeable interactions, which is due to the deceleration parameter $q$. Interaction terms given via Eq.~(\ref{eq:Q2}) are "non linear" interactions with fixed~($n=0$) and changeable~($n=1$) signs, respectively. To study the dynamics of the large scale universe, in Physical Literature a very high accurate approximation is used, namely, we consider a mixture of the dark energy and dark matter as the content of the universe. Which does mean that the effective fluid is described as
\begin{equation}\label{eq:rhoeff}
\rho_{eff} = \rho_{DM} + \rho_{DE},
\end{equation}
and 
\begin{equation}\label{eq:Peff}
P_{eff} = P_{DM} + P_{DE}.
\end{equation}
Therefore, with different assumptions about the dark energy and dark matter we are able to reconstruct the dynamics of the universe. Additional assumptions allowing us to construct various cosmological models we will present in section ~\ref{sec:IntModels}.
\\\\
The paper is organised as follows: In section~\ref{sec:IntModels} we will give the definition of the interacting models in modern cosmology and will present basics on the phase space analysis to find late time attractor solutions of the field equations of General Relativity. In section~\ref{sec:PSA} phase space analysis is performed, late time attractors are found and classified according to their cosmological applicability. To save a place we presented only attractor solutions. Finally, discussion on obtained results are summarised in section~\ref{sec:Discussion}.

\section{Interacting models and autonomous system}\label{sec:IntModels}
It is well known that to describe the dynamics of the large scale flat FRW universe we need the following set of equations obtained from the field equations of General Relativity
\begin{equation}\label{eq:Fridmman vlambda}
H^{2}=\frac{\dot{a}^{2}}{a^{2}}=\frac{8\pi G\rho}{3},
\end{equation}
\begin{equation}\label{eq:fridman2}
\frac{\ddot{a}}{a}=-\frac{4\pi G}{3}(\rho+3P).
\end{equation}
We suppose, that the cosmological constant $\Lambda=0$, the gravitational constant $G$ and $c$ are constants with $c=8\pi G=1$. On of the approaches to alleviate the cosmological coincidence problem is to consider interacting dark energy models. In modern cosmology an interaction between fluid components mathematically does mean~\cite{Gua}~-~\cite{Khurshudyan4}
\begin{equation}\label{eq:firstfluid}
\dot{\rho}_{m}+3H(\rho_{m}+P_{m})=Q,
\end{equation}
and
\begin{equation}\label{eq:secondfluid}
\dot{\rho}_{\Lambda}+3H(\rho_{\Lambda}+P_{\Lambda})=-Q.
\end{equation}
The form of $Q$ is determined under phenomenological assumptions, mainly, the dimensional analysis is used to construct interactions. It is reasonable to consider interactions which could improve previously known results and at the same time will not make the mathematical treatment of the problems complicated. It is widely believed that deeper understanding of the nature of dark energy and dark matter could give fundamental explanations of the phenomenological assumptions about interaction. It is known that a phase space of a dynamical system it is a space in which all possible states of the system are represented. To analyse the dynamical system of interacting Chaplygin gas we set~\cite{Huang}
\begin{equation}\label{eq:x}
x = \frac{\rho_{c}}{3H^{2}},
\end{equation}
\begin{equation}\label{eq:y}
y = \frac{P_{c}}{3H^{2}},
\end{equation}
\begin{equation}\label{eq:z}
z = \frac{\rho_{m}}{3H^{2}},
\end{equation}
and 
\begin{equation}\label{eq:N}
N=\ln{a},
\end{equation}
where $a$ it is the scale factor. To have physically reasonable solutions we should have the following constraints $0 \leq x \leq 1$ and $0 \leq z \leq 1$. At the same time we should remember that $x$ and $z$ according to Eq.~(\ref{eq:Fridmman vlambda}) should satisfy to the following constraint
\begin{equation}\label{eq:FConst}
x+z =1.
\end{equation}
In terms of $x$ and $y$ EoS parameter of the Chaplygin gas reads as
\begin{equation}\label{eq:EoSC}
\omega_{c} = \frac{y}{x},
\end{equation}
while the EoS parameter of the effective fluid reads as
\begin{equation}\label{eq:EoSeff}
\omega_{eff} = \frac{P_{c}}{\rho_{c} + \rho_{m}} = y,
\end{equation}
because dark matter is considered as a pressureless fluid. It is not hard to show that the deceleration parameter $q$ reads as
\begin{equation}\label{eq:q}
q = -1 - \frac{\dot{H}}{H^{2}} = \frac{1}{2}(1 + 3 y).
\end{equation}
There is a huge number of articles presenting phase space analysis of different cosmological models~\cite{Dagoberto}~-~\cite{Rong}~(to mention a few). From the next section we will start our study taking into account  general algorithm of finding critical points of the autonomous system $x^{\prime}$ and $y^{\prime}$, where $\prime$ it is the derivative with respect to $N$. Solutions of $x^{\prime}=0$ and $y^{\prime}=0$ should be found first, then the sign of the determinant and trace of the Jacobian matrix of $x^{\prime}$ and $y^{\prime}$ will point out the stability of the critical point. It is well known that if the trace of the Jacobian matrix is negative, while the determinate is positive, then the critical point is stable, because the real parts of eigenvalues are positive. This is in case of linear stability. From the other hand a stable critical point it is an attractor, which is we are looking for. Therefore, we need to find the range of the model parameters such, that to have a physically reasonable stable critical points i.e.  $0 \leq x_{c} \leq 1$ and $0 \leq z_{c} \leq 1$. 

\section{Phase space analysis}\label{sec:PSA}
In this section we will find late time attractors of the field equations for various interacting Chaplygin gas cosmological scenarios. For that we need to determine the form of the interaction term $Q$. Four different types of interaction are considered. Combining general experience known from Physical Literature we will impose the following constraints on the model parameters
\begin{equation}\label{eq:alpha}
0 < \alpha \leq 1,
\end{equation}
\begin{equation}\label{eq:b}
0\leq b < 1,
\end{equation}
\begin{equation}\label{eq:gamma}
0\leq \gamma < 1,
\end{equation}
and 
\begin{equation}\label{eq:A}
A\geq 0~~~~and~~~~B>0.
\end{equation}
In our calculations we used a fact that 
\begin{equation}
\rho^{\alpha} = \frac{B}{3H^{2} (Ax - y)}.
\end{equation}

\subsection{Interaction $Q = 3 H b \rho + \gamma \dot{\rho }$}\label{ssec:Q1}
\begin{figure}[h!]
 \begin{center}$
 \begin{array}{cccc}
\includegraphics[width=80 mm]{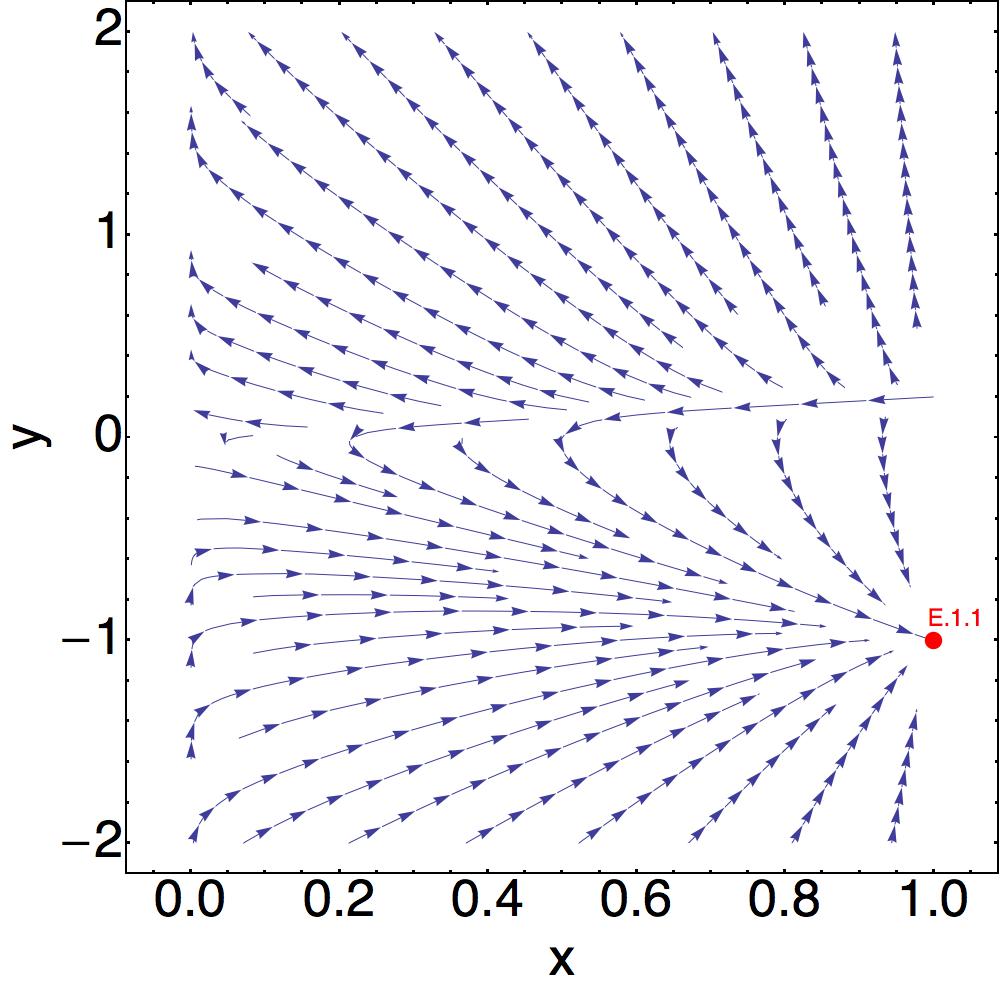}  &
\includegraphics[width=80 mm]{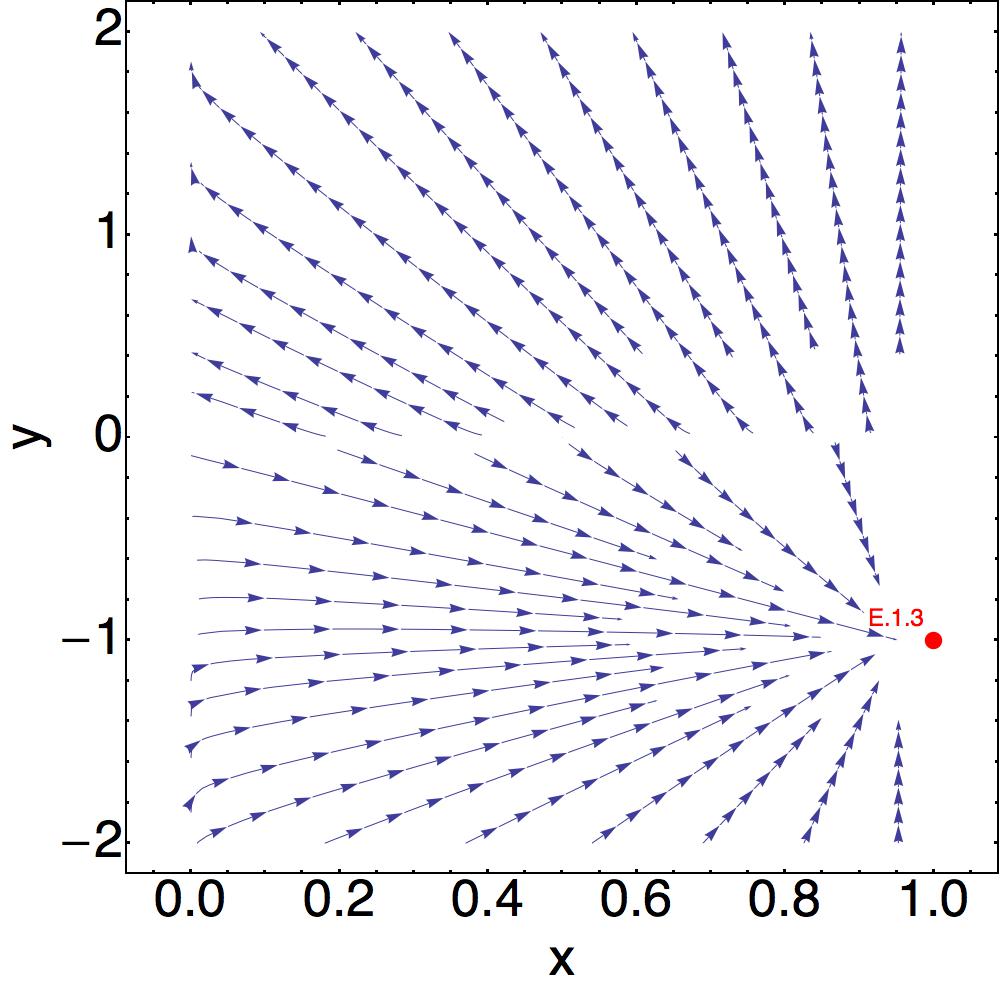}\\

 \end{array}$
 \end{center}
\caption{Phase space portraits for the models with $Q=3Hb\rho_{m} + \gamma \dot{\rho}_{m}$ and $Q=3Hb\rho_{c} + \gamma \dot{\rho}_{m}$. }
 \label{fig:Fig1}
\end{figure}

Interactions considered in this subsection are very well know and have been considered intensively. Our interest is to obtain the late time attractors of the field equations where the interactions are particular examples obtained from the following interaction 
\begin{equation}\label{eq:Q1GenInit}
Q = 3 H b  \rho  + \gamma \dot{\rho},
\end{equation}
where $\rho$ could be either the energy density of the effective fluid or the energy density one of the components of the effective fluid. The general form of the interaction term given via Eq.~(\ref{eq:Q1GenInit}) for two fluid universe reads as
\begin{equation}\label{eq:Q1GenF}
Q = 3 H b (\rho_{m} + \rho_{c}) + \gamma (\dot{\rho}_{m} + \dot{\rho}_{c}),
\end{equation}
which does allow us to obtain an explicit form of the autonomous system
\begin{equation}\label{eq:xpQ1}
x^{\prime} = \frac{dx}{dN} = -3 \left ( b -\gamma -  y (-1+x+y) \right ),
\end{equation}
and 
\begin{equation}\label{eq:ypQ1}
y^{\prime} = \frac{dy}{dN} = 3y(1+y) - 3  \frac{\left ( -\alpha y + A(1+\alpha) x \right ) \left (  b+ x+ y - \gamma (1+y) \right )}{x}.
\end{equation}
To find stable critical points~(attractors), we need to solve $x^{\prime} = 0$ and $y^{\prime}=0$ equations and determine the signs of the eigenvalues of the appropriate Jacobian matrix of  $x^{\prime}$ and $y^{\prime}$. Table \ref{tab:Table1} does represent late time attractors for different forms of interactions obtained from Eq.~(\ref{eq:Q1GenInit}). $E.1.1$ is a stable attractor for $\alpha$, $b$, $\gamma$ and $A$ satisfying to Eq.-s~(\ref{eq:alpha})~-~(\ref{eq:A}). $E.1.1$ does represent Chaplygin gas dominating solution where $\omega_{c} =-1$, $\omega_{eff} = -1$ and $q=-1$ i.e. $\Lambda$CDM model is recovered. Stable attractor $E.1.2$ is a scaling solution since
\begin{equation}
r = \frac{\Omega_{m}}{\Omega_{c}} = b,
\end{equation}
when $b\neq 0$. $\omega_{c} = -1-b$ does mean that phantom phase is on face in this case. Attractor $E.1.3$ is of the same nature as $E.1.2$, while attractor $E.1.4$ has the same nature as $E.1.1$. Late time attractor $E.1.5$ has completely different nature, however in this case an accelerated expansion is not possible. $E.1.6$, $E.1.7$ and $E.1.8$ late time attractors are scaling solutions~($b \neq 0$) since
\begin{equation}
r = \frac{\Omega_{m}}{\Omega_{c}} = \frac{b}{1-b}.
\end{equation}
In this case according to Eq.(\ref{eq:q}) $q =-1$, while according to Eq.~(\ref{eq:EoSC}) 
\begin{equation}
\omega_{c} = -\frac{1}{1-b}, 
\end{equation}
EoS parameter of the effective fluid according to Eq.~(\ref{eq:EoSeff}) is equal to $-1$. In Fig.~(\ref{fig:Fig1}) we have illustrated phase space portraits corresponding to $E.1.1$ and $E.1.3$.

\begin{table}
  \centering
    \begin{tabular}{ | l | l | l | l | p{2cm} |}
    \hline
  $Q$ & $S. P.$ & $x$ & $y$ \\
  \hline
    $3Hb\rho_{m} + \gamma \dot{\rho}_{m}$ & $E.1.1$  & $1$ & $-1$ \\
     \hline
      $3Hb \rho_{c} + \gamma \dot{\rho}_{c}$ & $E.1.2$  & $1/(1+b)$  & $-1$ \\
          \hline
     $3Hb\rho_{c} + \gamma \dot{\rho}_{m}$ &$E.1.3$  &  $1/(1+b)$ & $-1$ \\
    \hline
      $3Hb\rho_{m} + \gamma \dot{\rho}_{c}$ &$E.1.4$  &  $1$ & $-1$ \\
    \hline
      $3Hb\rho_{m} + \gamma \dot{\rho}_{c}$ &$E.1.5$  &  $\frac{A-b - \gamma - r}{2A(1+\gamma)}$ & $-\frac{b+\gamma - A + r}{2(1+\gamma)}$ \\
    \hline
      $3Hb\rho + \gamma \dot{\rho}$ &$E.1.6$  &  $1-b$ & $-1$ \\
    \hline
        $3Hb(\rho_{c} +\rho_{m})$ &$E.1.7$  &  $1-b$ & $-1$ \\
    \hline
       $\gamma \dot{\rho}$ &$E.1.8$  &  $1-b$ & $-1$ \\
    \hline

    \end{tabular}
\caption{ Critical points corresponding to various interaction terms obtained from general form Eq.~(\ref{eq:Q1}) for $n=0$. $r=\sqrt{4Ab(1+\gamma) + (b-A+\gamma)^{2}}$.}
  \label{tab:Table1}
\end{table}

\subsection{Interaction $Q = q(3Hb \rho  + \gamma \dot{\rho} )$}\label{sseq:Q2}
\begin{figure}[h!]
 \begin{center}$
 \begin{array}{cccc}
\includegraphics[width=80 mm]{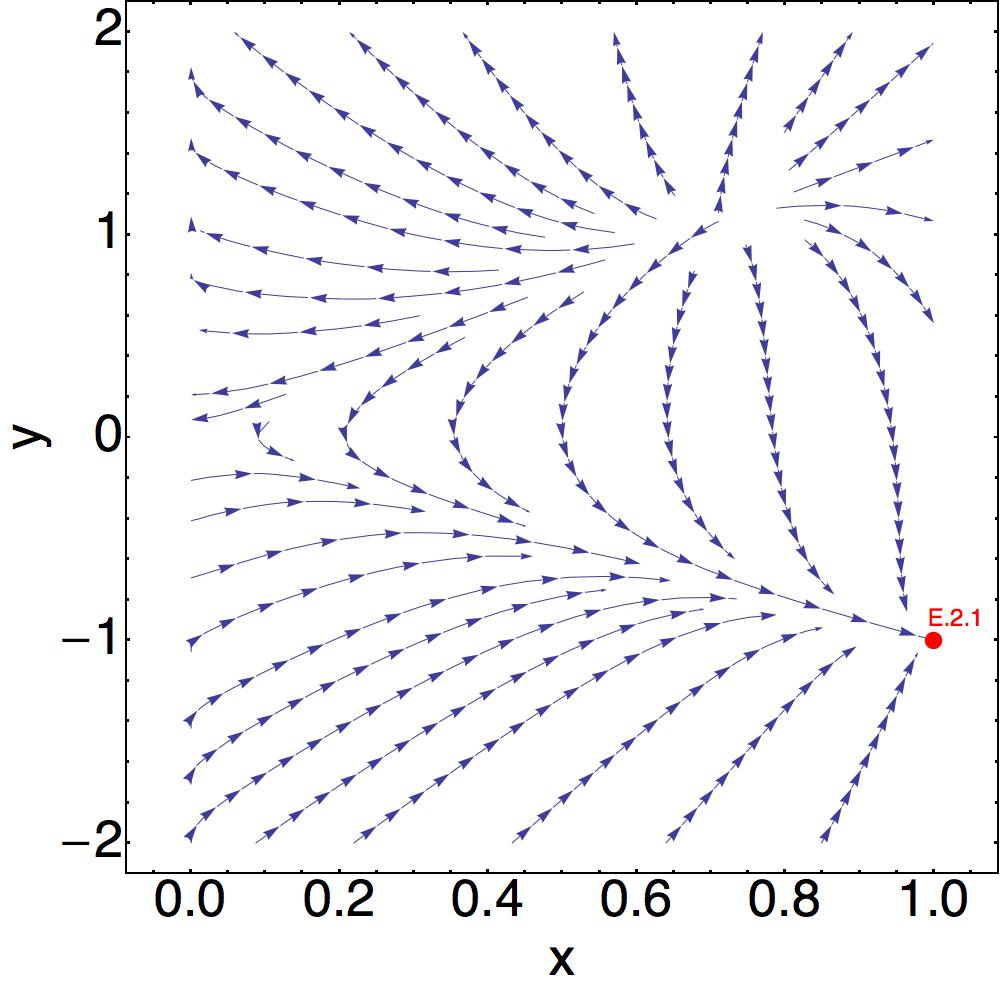}  &
\includegraphics[width=80 mm]{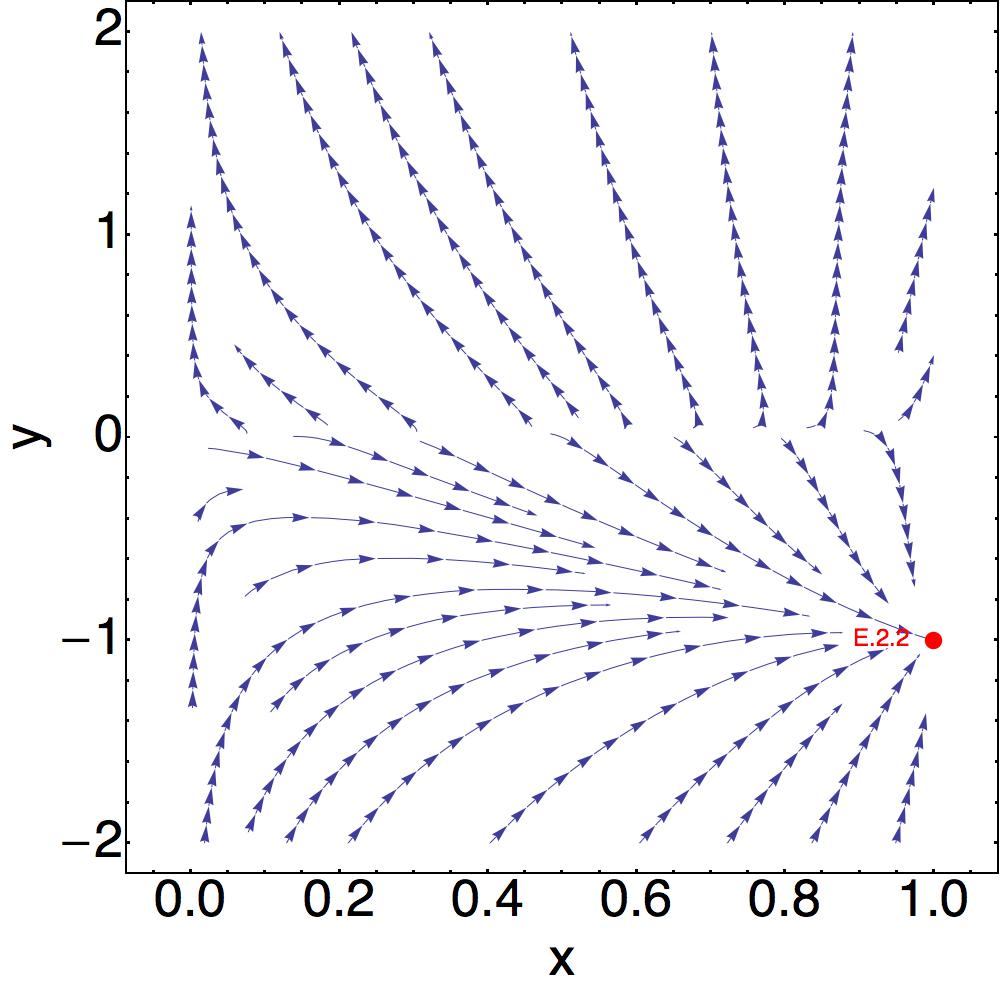}\\

 \end{array}$
 \end{center}
\caption{Phase space portraits for the models with $Q= q (3Hb\rho_{c} + \gamma \dot{\rho}_{c} )$ and $Q=\gamma q ( \dot{\rho}_{c} +  \dot{\rho}_{m})$. }
 \label{fig:Fig2}
\end{figure}

The second class of interactions that we will consider could be obtained from the following general form~(two fluid case)
\begin{equation}\label{eq:Q2GenF}
Q = q \left ( 3 H b ( \rho_{c} + \rho_{m}) + \gamma (\dot{\rho}_{c} + \dot{\rho}_{c} ) \right ).
\end{equation}
It is know as sign changeable interaction, where the deceleration parameter $q$ is used to have the desirable effect. For the general case given via Eq.~(\ref{eq:Q2GenF}), the critical points are solutions of the following two equations
\begin{equation}\label{eq:xprimeQ2}
x^{\prime} = \frac{3}{2} \left ( -b(1+3y) +\gamma + y (-2+2x+4\gamma +3 \gamma y) \right ),
\end{equation}
and
\begin{equation}\label{eq:yprimeQ2} 
y^{\prime} = 3y (1+y) + \frac{3}{2x} \left ( -\alpha y + A(1+\alpha) x\right) \left( b (1+3y) + 2 (x+y) - (1+y) (1+3y) \gamma \right).
\end{equation}
Late time attractors for different interactions obtained from Eq.~(\ref{eq:Q2GenF}) are collected in Table~\ref{tab:Table2}. Our analysis does show that late attractors are possible to obtain if $b =0$~(Fig.~(\ref{fig:Fig2})).
\begin{table}
  \centering
    \begin{tabular}{ | l | l | l | p{2cm} |}
    \hline
  $Q$ & $S. P.$ & $x$ & $y$ \\
  \hline
    $q( 3Hb\rho_{c} + \gamma \dot{\rho}_{c})$ & $E.2.1$  & $1/(1-b)$ & $-1$ \\
     \hline
      $\gamma q ( \dot{\rho}_{c} + \dot{\rho}_{m})$ & $E.2.2$  & $1$  & $-1$ \\
          \hline
     $b q (3H b (\rho_{c}+\rho_{m}) + \dot{\rho}_{m} +\dot{\rho}_{c})$ &$E.2.3$  &  $1+b$ & $-1$ \\
    \hline
      $3Hb q ( \rho_{m} + \rho_{c} )$ &$E.2.4$  &  $1+b$ & $-1$ \\
    \hline
      $3Hb q \rho_{c} $ &$E.2.5$  &  $1/(1-b)$ & $-1$ \\
    \hline
      $3Hb q \rho_{m}$ &$E.2.6$  &  $1$ & $-1$ \\
    \hline
           \end{tabular}
\caption{ Critical points corresponding to various interaction terms obtained from general form Eq.~(\ref{eq:Q1}) for $n=1$.}
  \label{tab:Table2}
\end{table}

\subsection{Interaction $Q = 3 H b \left( \rho + \frac{\rho_{i}\rho_{j}}{ \rho} \right )$} \label{ssec:Q3}
\begin{figure}[h!]
 \begin{center}$
 \begin{array}{cccc}
\includegraphics[width=80 mm]{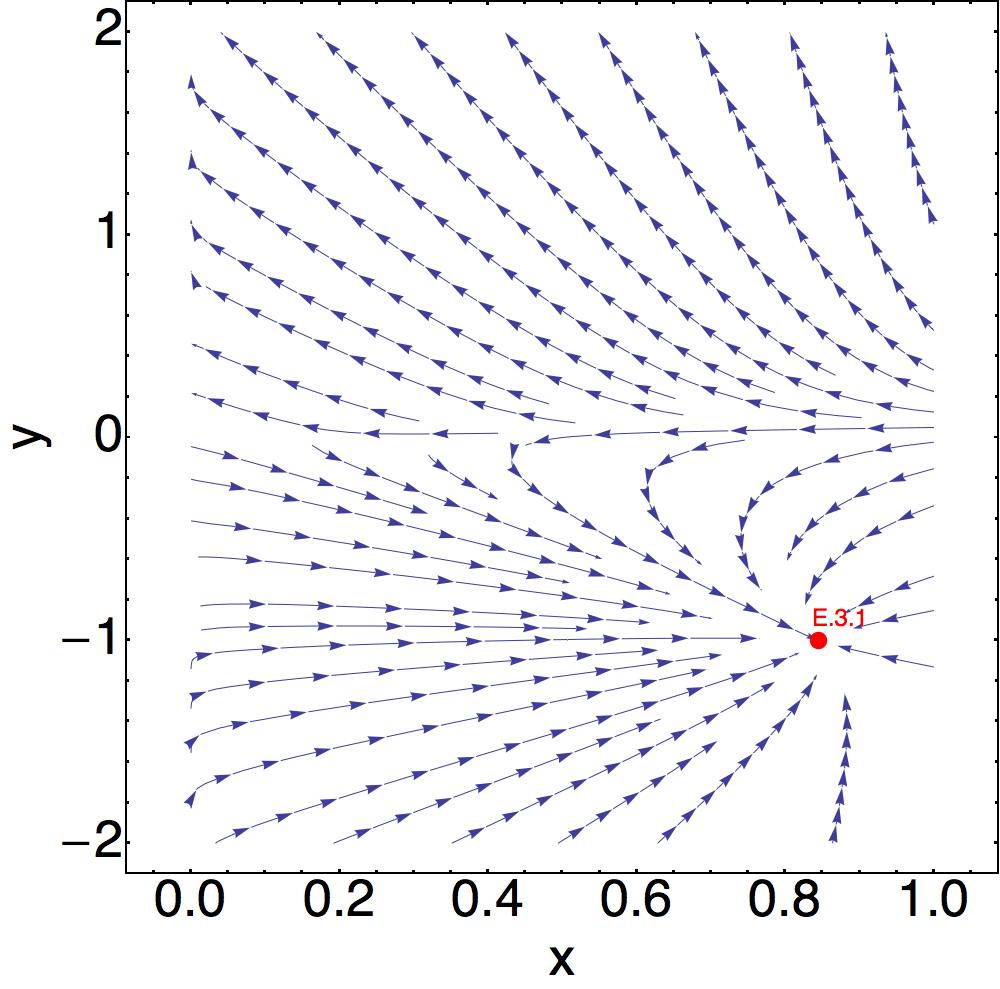}  &
\includegraphics[width=80 mm]{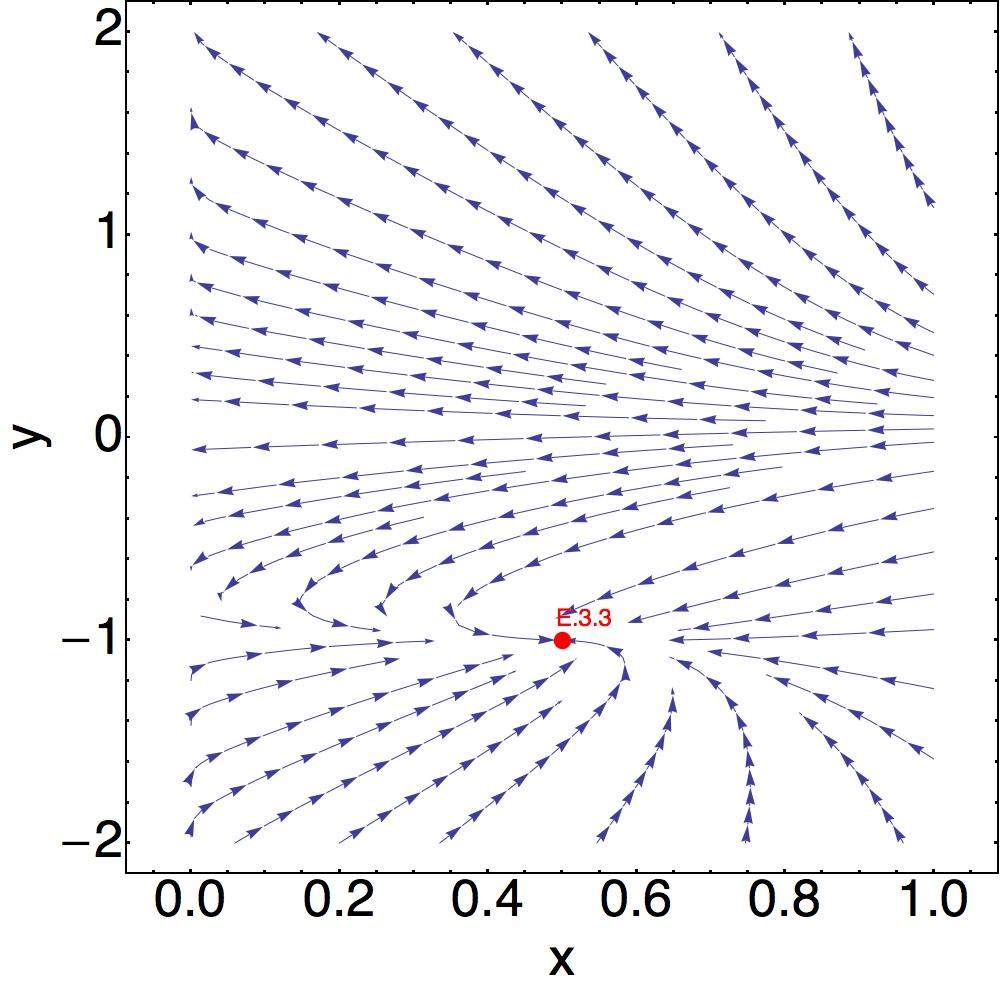}\\

 \end{array}$
 \end{center}
\caption{Phase space portraits for the models with interactions $3Hb \left ( \rho_{c}  + \rho_{c}^{2}/(\rho_{c}+ \rho_{m})\right)$ and  $3Hb \left ( \rho_{c}  + \rho_{m} + \rho_{c}^{2}/(\rho_{c}+ \rho_{m})\right)$. }
 \label{fig:Fig3}
\end{figure}

In this subsection we will consider a non linear interaction. In Physical Literature there is an active discussion on non linear interactions~(see for instance~\cite{Yin},~\cite{Arevalo}). The non linear interaction forms considered in this paper have a phenomenological origin and are based on the dimensional analysis. The general form of the non linear interaction considered by us has the following form~(for two fluid models)
\begin{equation}\label{eq:Q3GenF}
Q = 3 H b \left ( \rho_{c} +  \rho_{m}  + \frac{\rho_{ij}} {\rho_{c} + \rho_{m}} \right ),
\end{equation}
where $\rho_{ij}$ it is the product of $\rho_{c}$ and $\rho_{m}$ and if $i = j = m$ we have $\rho_{m}^{2}$ etc. In Table~\ref{tab:Table3} we summarize late time attractors which we have obtained using different forms of interaction term $Q$ having similar forms with Eq.~(\ref{eq:Q2}) for $n=0$. $E.3.1$ is a scaling late time attractor for $0 < \alpha \leq 1$, $0<b<1$ and $A\geq 0$, because in this case
\begin{equation}
\frac{\Omega_{m}}{\Omega_{c}} = \frac{1}{2} \left ( b -1 + \sqrt{1+b (6+b)}\right ).
\end{equation}
It is not hard to see that $\omega_{eff} = -1$, $q=-1$ and
\begin{equation}
\omega_{c} = \frac{2b}{1+b - \sqrt{1+b(6+b)}}.
\end{equation}
Another scaling late time attractor is possible to obtain when $0 < b < 1/2$,  $0 <  \alpha \leq 1$ and $A\geq 0$, if we consider the following interaction term $Q$
\begin{equation}
Q= 3Hb \left ( \rho_{c}+\rho_{m} + \frac{\rho_{m}^{2}} {\rho_{c} + \rho_{m}}  \right ).
\end{equation}
In Table~\ref{tab:Table3} it does correspond to $E.3.2$, for which
\begin{equation}
\frac{\Omega_{m}}{\Omega_{c}} = -\frac{1}{2} + \frac{\sqrt{1-4b^{2}}} {2(1-2b)}.
\end{equation}
It is not hard to see that
\begin{equation}
\omega_{c} = - \frac{2b}{2b-1 + \sqrt{1-4b^{2}}}.
\end{equation}
Another scaling late time attractor is $E.3.3$, when  $0 < \alpha \leq 1$, $0<b<1$ and $A\geq 0$ with
\begin{equation}
\frac{\Omega_{m}}{\Omega_{c}} = \frac{2b - 1 + \sqrt {1-4(b-1)b}}{2(b-1)},
\end{equation}
\begin{equation}
\omega_{c} = - \frac{2b}{ \sqrt {1-4(b-1)b} - 1},
\end{equation}
and $q=-1$, $\omega_{eff} =-1$. Finally, if we consider interaction of the following form
\begin{equation}
Q = 3 H b \left ( \rho_{m} + \frac{\rho_{m}^{2}}{\rho_{c} + \rho_{m}} \right ),  
\end{equation}
then late time attractor $E.3.4$ could be found, which describes Chaplygin gas dominating phase with $\omega_{c} = -1$. In all cases an accelerated expansion is on face. Consideration of non linear interactions does give us qualitatively four different scaling solutions~(among considered interactions). Therefore, future consideration of additional possibilities will enrich our experience.  In Fig.~(\ref{fig:Fig3}) we have illustrated phase space portraits corresponding to $E.3.1$ and $E.3.3$.

\begin{table}
  \centering
    \begin{tabular}{ | l | l | l | p{2cm} |}
    \hline
  $Q$ & $S. P.$ & $x$ & $y$ \\
  \hline
    $ 3Hb \left ( \rho_{c}  + \rho_{c}^{2}/(\rho_{c}+ \rho_{m})\right)$ & $E.3.1$  & $(-1-b + r_{1})/2b$ & $-1$ \\
     \hline
      $ 3Hb \left ( \rho_{c}  + \rho_{m} + \rho_{m}^{2}/(\rho_{c}+ \rho_{m})\right)$ & $E.3.2$  & $(-1+2b + r_{2})/2b$  & $-1$ \\
          \hline
     $3Hb \left ( \rho_{c}  + \rho_{m} + \rho_{c}^{2}/(\rho_{c}+ \rho_{m})\right)$ &$E.3.3$  &  $(r_{3} - 1)/3b$ & $-1$ \\
    \hline
    $ 3Hb \left ( \rho_{m}  + \rho_{m}^{2}/(\rho_{c}+ \rho_{m})\right)$ & $E.3.4$  & $1$ & $-1$ \\
     \hline

               \end{tabular}
\caption{ Critical points corresponding to various interaction terms obtained from general form Eq.~(\ref{eq:Q2}) for $n=0$, where $r_{1} = \sqrt{1+b (6+b)}$,$r_{2} = \sqrt{1-4b^{2}}$ and $r_{3} = \sqrt {1-4(b-1)b}$. }
  \label{tab:Table3}
\end{table}

\subsection{Interaction $Q = 3 H b q \left( \rho + \frac{\rho_{i}\rho_{j}}{ \rho} \right )$} \label{ssec:Q4}
Consideration of non linear sign changeable interactions is another phenomenological assumption of this work. Mainly, motivated by the results obtained in subsection~\ref{ssec:Q3}, we decided to consider the following possibility
\begin{equation}\label{eq:Q4GenF}
Q = 3 H b q \left ( \rho_{c} +  \rho_{m}  + \frac{\rho_{ij}} {\rho_{c} + \rho_{m}} \right ).
\end{equation}
We followed to the receipts of the previous subsections to find late time attractors for this case as well. Without going into all details of our calculations, we would like to report that  we have considered two particular forms for the interaction term $Q$ and found that late time attractors do not exist. Considered interaction terms are
\begin{equation}
Q = 3 H b q \left ( \rho_{c} + \frac{\rho_{c}^{2}}{\rho_{c} + \rho_{m}}\right ),
\end{equation}  
and
\begin{equation}
Q = 3 H b q \left ( \rho_{m} + \frac{\rho_{m}^{2}}{\rho_{c} + \rho_{m}}\right ).
\end{equation}  
We are not closing our study on this type of sign changeable interactions and we will come back to them in our next works. 
 
\section{\large{Discussion}}\label{sec:Discussion}
Subject of our interest has been a study of interacting Chaplygin gas cosmological models. Besides, we had a goal to construct some sign changeable non linear interactions. Instead to solve field equations for some initial conditions, we performed phase space analysis. Adopting appropriate notations allowed us to obtain an appropriate autonomous system and start a study on the stability of the critical points. Appropriate restrictions from the cosmological and astrophysical studies know in Physical Literature are imposed on some parameters of the models to find physically reasonable solutions. Four types of interaction are considered. Two of them are fixed sing and sign changeable non linear interactions. For each case late time attractors are found. Our study showed that in case of some of the sign changeable non linear interactions, constructed in this paper, stable critical points are missing. While a study of the models when $Q = q(3Hb \rho + \gamma \dot{\rho})$ reveals that only one type late time attractor is possible, namely  $x=1$ and $y=-1$ describing Chaplygin gas dominating phase with $\omega_{c} =-1$ and $q=-1$. The situation is completely different if we consider fixed sign interactions. In case of non linear interactions~(among considered ones) four different types of late time scaling solutions were found, while for the models with $Q = 3Hb \rho + \gamma \dot{\rho}$ only two late time scaling solutions are found. One of them is described the universe in a phantom regime. More general cases of interacting models could be studied numerically and we hope to report some new results in our next articles.  

\section*{Acknowledgement}
M. Khurshudyan is grateful to Prof E. Kokanyan for a warm hospitality, comprehensive support and stimulating discussion.

\newpage

\end{document}